\documentclass[preprint,aps,prb]{revtex4}
\usepackage{amssymb}
\usepackage{graphicx}
\usepackage{dcolumn}
\usepackage{latexsym,bm}

\begin{document}

\begin{titlepage}

\setcounter{page}{1} \pagenumbering{arabic}
\title{Polarization Induced Switching Effect in Graphene
Nanoribbon Edge-Defect Junction}

\author{G. Yin}
\affiliation{Department of Physics, Fudan University, Shanghai
200433, People's Republic of China}

\author{Y. Y. Liang}
\affiliation{Department of Physics, Fudan University, Shanghai
200433, People's Republic of China}

\author{F. Jiang}
\affiliation{Department of Physics, Fudan University, Shanghai
200433, People's Republic of China}

\author{H. Chen\footnote{Author to whom correspondence should be addressed.
Electronic mail: haochen@fudan.edu.cn}} \affiliation{Department of
Physics, Fudan University, Shanghai 200433, People's Republic of
China}

\author{P. Wang}
\affiliation{High-end computing center, Fudan University, Shanghai
200433, People's Republic of China}

\author{R. Note}
\affiliation{Institute for Materials Research, Tohoku University,
Sendai 980-8577, Japan}

\author{H. Mizuseki}
\affiliation{Institute for Materials Research, Tohoku University,
Sendai 980-8577, Japan}

\author{Y. Kawazoe}
\affiliation{Institute for Materials Research, Tohoku University,
Sendai 980-8577, Japan}

\date{\today}

\begin{abstract}
With nonequilibrium Green's function approach combined with density
functional theory, we perform an ab initio calculation to
investigate transport properties of graphene nanoribbon junctions
self-consistently. Tight-binding approximation is applied to model
the zigzag graphene nanoribbon (ZGNR) electrodes, and its validity
is confirmed by comparison with GAUSSIAN03 PBC calculation of the
same system. The origin of abnormal jump points usually appearing in
the transmission spectrum is explained with the detailed
tight-binding ZGNR band structure. Transport property of an edge
defect ZGNR junction is investigated, and the tunable tunneling
current can be sensitively controlled by transverse electric fields.
\end{abstract}

\maketitle

\end{titlepage}

\section{INTRODUCTION}

With the stunning developments of electronic engineering technology,
transistor density of silicon based semiconductor chips is approaching the
ultimate size limitation. Innovative materials such as organic molecules\cite%
{Pantelides M,Reed3,Reed2,Reed,Jiang2,Liang1,Liang2,H Chen,YX Zhou} and
single-walled carbon nanotubes\cite{Baughman} (SWNTs) have been suggested as
substitutions of conventional semiconductors to implement promising
electronic devices in nanoscale where quantum mechanics dominates the
electron kinetic behavior.\cite{Datta:Electronic Transport in Mesoscopic
Systems} Graphene, the single layer of carbon honeycomb two-dimensional ($2$%
D) crystal structure, has become an active field of research because of its
exotic physical properties and outstanding electronic quality.\cite{A. K.
Geim,Yuanbo Zhang,K. S. Novoselov,K. S. Novoselov2} The strong C-C bonds in
the plane make graphene an easily available material. Experimental results
indicate that, electron mobility of graphene sheet, which is weakly
temperature dependent, can be up to $10^{2}$ cm$^{\text{2}}$V$^{\text{-1}}$s$%
^{\text{-1}}$ with the carrier density of $10^{12}$ cm$^{\text{-2}}$ at room
temperature.\cite{A. K. Geim} These exceptional properties make graphene a
promising candidate material for nanoscale electronic devices, and have
aroused considerable attention in both academic and industrial worlds.\cite%
{IEEE,Westervelt}

Graphene nanoribbons (GNRs) are quasi one-dimensional (1D) structures cut
from graphene sheet in particular orientations. They are categorized by the
alignment of the edge atoms, namely, zigzag (ZGNR) and armchair (AGNR) ones.%
\cite{Kyoko Nakada} With state-of-the-art experimental technology such as
direct focused electron writing\cite{focus electron} or plasma etching\cite%
{M. Y. Han}, structure modulation of GNR is already possible. Recent
first-principles calculation indicates that, in contrast with SWNTs, both
ZGNRs and AGNRs are semiconducting, exhibiting band gap increase with the
decrease in the ribbon width.\cite{Barone,Young-Woo Son,Z. Chen,M. Y. Han}
Extracting ultra smooth GNRs from solution--derived graphite, Wang and Li
\emph{et al.} composed sub-$10$ nm GNR based FETs with considerable on/off
ratio due to the semiconducting property of narrow GNRs.\cite{Xinran
Wang,Xiaolin Li}

To investigate transport properties of nanoscale systems, nonequilibrium
Green's function (NEGF) technique combined with density functional theory
(DFT) has been developed into a standard and promising method,\cite{Y.
Xue,Datta2,J. Taylor,S. Datta,Taylor2} which is also widely used in GNR
device designing.\cite{B DOPING,Z,symmetry,IEEE SYM,HP Cheng,FET
NEGF,Different Electrodes,defect NEGF,Z2} In this scheme, Liang \emph{et al.}%
\cite{Different Electrodes} compared several graphene based FET contacts,
and suggested ZGNRs to be the best choice. With ZGNR source and drain
electrodes, Yan \emph{et al.}\cite{Z} designed a Z shaped GNR-FET with
on/off ratio up to $10^{4}$, in which an AGNR is applied in the middle to
function as the semiconducting channel. Defects in GNR structures, such as
vacancy or doped atoms, are also reported to modify GNR electronic
properties significantly\cite{DFCT Structure,defect NEGF}.

In this paper, we apply DFT-NEGF method to simulate GNR based nanoelectronic
devices, and a detailed investigation of GNR junctions' transport properties
is carried out. Particular attention is paid to the external electrostatic
field controlled switching effects. This paper is organized as follows: In
Sec. II, a detailed description of numerical methods and theoretical
formulism is presented. Calculated transport properties of GNR devices and
the physical origins are given in Sec. III. We summarize our results in Sec.
IV.

\section{THEORETICAL FORMULAS AND CALCULATION DETAILS}

In the calculation, the surface Green's functions of ZGNR electrodes are
calculated self-consistently under the DFT based tight binding approximation
(TB), while the self-consistent DFT-NEGF transport calculation is applied to
the scattering region attached by the left and right electrodes. Both the
DFT-TB and DFT-NEGF computations are implemented based on quantum chemistry
software GAMESS(US),\cite{Gamess1,M.S.Gordon} while structure optimization
and electron structure calculation under periodic boundary conditions (PBC)
are carried out by GAUSSIAN03.\cite{M.J. Frisch}

\begin{figure}[tbp]
\includegraphics[width=1.0\textwidth]{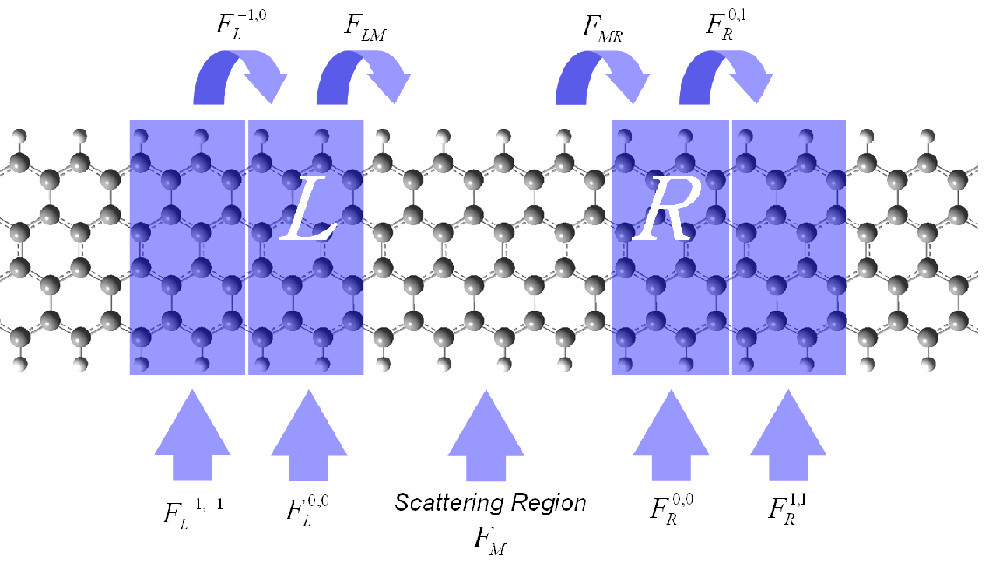}
\caption{G. Yin \emph{et al.}}
\label{alignment}
\end{figure}

In order to deal with the contact-device-contact open system, semi-infinite
ZGNR electrodes and the central scattering region are considered separately.
The overall Hamiltonian of the entire open system in Eq.(\ref{MATRIX}) is
theoretically an infinite matrix. Atom layers corresponding to the
submatrices in Eq.(\ref{MATRIX}) are denoted in Fig.\ref{alignment}, where
the source and drain electrodes are composed of two semi-infinite ZGNRs with
$4$ zigzag carbon chains (4-ZGNR). The structure is fully relaxed by DFT
geometry optimization under PBC, which is carried out by GAUSSIAN03 under
GTO basis 6-31G and B3LYP exchange-correlation term.\cite{B3,LYP} As the
starting point of geometry optimization, the C-C and C-H bond lengths of a
unit cell are set at $1.426$ \AA\ and $1.070$ \AA\ respectively, and the
lattice constant of the perfect quasi-1D crystal is finally relaxed to $%
4.9302$ \AA . $F_{M}$ is the Fock submatrix of the middle scattering region,
while $F_{L}^{0,0}$ and $F_{R}^{0,0}$ are the submatrices of the contact
"surface" layers of the left and right electrodes respectively. In our NEGF
transport simulation, semi-infinite leads are simulated under TB regime, in
which only the interactions between the closest TB unit cells are taken into
account ($F_{L}^{-1,0},F_{R}^{0,1}$). It is lately reported that a TB unit
cell of ZGNR should at least include two armchair carbon atom chains to
avoid crucial inter-cell interacting information loss.\cite{HP Cheng}
\begin{equation}
\left[
\begin{array}{ccccccc}
\cdots & \cdots & \cdots & \cdots & \cdots & \cdots & \cdots \\
\cdots & F_{L}^{-1,-1} & F_{L}^{-1,0} & 0 & 0 & 0 & \cdots \\
\cdots & F_{L}^{0,-1} & F_{L}^{0,0} & F_{LM} & 0 & 0 & \cdots \\
\cdots & 0 & F_{ML} & F_{M} & F_{MR} & 0 & \cdots \\
\cdots & 0 & 0 & F_{RM} & F_{R}^{0,0} & F_{R}^{0,1} & \cdots \\
\cdots & 0 & 0 & 0 & F_{R}^{1,0} & F_{R}^{1,1} & \cdots \\
\cdots & \cdots & \cdots & \cdots & \cdots & \cdots & \cdots%
\end{array}%
\right]  \label{MATRIX}
\end{equation}

Combining the electrodes and the scattering region, the open system of GNR
junction can be handled as an isolated one with NEGF technique, and the
retarded Green's function of the scattering region is obtained with
\begin{equation}
G_{M}^{r}=\left( \epsilon ^{+}S_{M}-F_{M}-\Sigma _{L}^{r}-\Sigma
_{R}^{r}\right) ^{-1}  \label{GR}
\end{equation}%
where $\epsilon ^{+}$ denotes energy plus an infinitesimal imaginary part ($%
1\times 10^{-6}$), and $S_{M}$ is the overlap matrix due to the
nonorthogonality of the GTO basis set. Including the information of two
semi-infinite leads, retarded self energy $\Sigma _{L}^{r}$ and $\Sigma
_{R}^{r}$ are calculated from $g_{i}^{r}$ ($i=L,R$), the retarded surface
green's function (SGF) of the interface layer \emph{L} and \emph{R}.

\begin{equation}
\Sigma _{i}^{r}=\left( \epsilon ^{+}S_{Mi}-F_{Mi}\right) g_{i}^{r}\left(
\epsilon ^{+}S_{iM}-F_{iM}\right)  \label{SelfE}
\end{equation}

$F_{ML}$($F_{MR}$), $S_{ML}$($S_{MR}$) are Fock and overlap matrices given
by DFT computation of lead-junction-lead open system, describing the
interactions between the scattering region and the left (right) electrode.
SGFs of contact layers are obtained by transformation matrix method
self-consistently with Eq.(\ref{SGFL}) and Eq.(\ref{SGFR}) in real space.%
\cite{SGF}

\begin{equation}
g_{L}^{r}=\left\{ \epsilon ^{+}S_{L}^{0,0}-F_{L}^{0,0}-\left(
F_{L}^{0,-1}-\epsilon ^{+}S_{L}^{0,-1}\right) g_{L}^{r}\left[ \left(
F_{L}^{0,-1}\right) ^{\intercal }-\epsilon ^{+}\left( S_{L}^{0,-1}\right)
^{\intercal }\right] \right\} ^{-1}  \label{SGFL}
\end{equation}

\begin{equation}
g_{R}^{r}=\left\{ \epsilon ^{+}S_{R}^{0,0}-F_{R}^{0,0}-\left(
F_{R}^{0,1}-\epsilon ^{+}S_{R}^{0,1}\right) g_{R}^{r}\left[ \left(
F_{R}^{0,1}\right) ^{\intercal }-\epsilon ^{+}\left( S_{R}^{0,1}\right)
^{\intercal }\right] \right\} ^{-1}  \label{SGFR}
\end{equation}

Under TB approximation, only the on-site Fock matrix of the left (right)
contact layer $F_{L}^{0,0}$ ($F_{R}^{0,0}$) and the closest neighbor
interaction $F_{L}^{0,-1}$ ($F_{R}^{0,1}$) are involved in the SGF
self-consist calculation. With retarded Green's function of the scattering
region, the density matrix can be given as

\begin{equation}
\rho =\frac{1}{2\pi }\int_{-\infty }^{\infty }\left[ G_{M}^{r}\left(
f_{L}\Gamma _{L}+f_{R}\Gamma _{R}\right) G_{M}^{a}\right] d\epsilon
\label{density}
\end{equation}%
where $G_{M}^{a}=(G_{M}^{r})^{\dag }$ is the advanced Green's function of
the scattering region. $\Gamma _{L(R)}=i(\Sigma _{L(R)}^{r}-[\Sigma
_{L(R)}^{r}]^{\dagger })$ denotes the electrode-device coupling matrix; $%
f_{L(R)}(\epsilon )$=$1/(1+e^{(\epsilon -\mu _{L(R)})/k_{B}T})$ represents
Fermi distribution of the left and right lead respectively, in which $T$ is
the environment temperature of the junction, $\mu _{L(R)}=E_{f}\pm \frac{1}{2%
}eV$ is the chemical potential for source and drain, and $E_{f}$ stands for
Fermi level of the semi-infinite leads.

Because of charging effect and level broadening in the scattering region
caused by electrode-device coupling, density matrix $\rho $, Fock matrix $%
F_{M}$, and retarded Green's function $G_{M}^{r}$ must be calculated
self-consistently. First, the converged density matrix of isolated cluster
based DFT calculation is applied as the initial guess. Then, we use the open
system self-consistent iteration loop to replace the original cluster-based
DFT loop of GAMESS(US) to obtain the density matrix from Eq.(\ref{density}).
The loop will not be stopped until the updated density matrix meets the
convergence criteria.

With converged $G_{M}^{r}$, the total tunneling current of the open system
at certain temperature can be given by Landauer-B{\"{u}}tiker equation\cite%
{R. Landauer}

\begin{equation}
I=\frac{2e}{h}\int (f_{R}-f_{L})T\left( \epsilon \right) d\epsilon
\label{cur}
\end{equation}

\begin{equation}
T\left( \epsilon \right) =\mbox{Tr}\left( \Gamma _{L}G_{M}^{r}\Gamma
_{R}G_{M}^{a}\right)  \label{Transmission}
\end{equation}

\noindent where $T\left( \epsilon \right) $ is the transmission function.

From the NEGF approach, the infinite open system can be solved as an
isolated one. In this work, DFT-NEGF computation is performed under
real-space GTO with the smallest effective core potential CEP-4G.\cite%
{CEP-4G} Becke-3 hybrid functional and Perdew-Wang-91 gradient-corrected
correlation functional\cite{B3,PW91} are used as the exchange-correlation
part.

To guarantee the validity of the TB approximation for Eq.(\ref{SGFL}) and
Eq.(\ref{SGFR}), we compare the TB band structure $E(kD)$ for the perfect
4-ZGNR with a periodical boundary condition (PBC) result in GAUSSIAN03,
where all inter-cell couplings are included.

Since GTO basis functions are originally denoted in real space, to calculate
the band structure of an infinite 1D bulk crystal, one need to transform the
Fock and overlap matrices to reciprocal space,
\begin{equation}
F_{mn}^{\vec{k}}=\left\langle \Psi _{m}^{\vec{k}}\left\vert \hat{F}%
\right\vert \Psi _{n}^{\vec{k}}\right\rangle =\sum_{i}e^{i\vec{k}\cdot \vec{R%
}_{i}}\left\langle \Psi _{m}^{0}\left\vert \hat{F}\right\vert \Psi _{n}^{%
\vec{R}_{i}}\right\rangle =\sum_{i}e^{i\vec{k}\cdot \vec{R}_{i}}F_{mn}^{0,%
\vec{R}_{i}},  \label{Fmnk}
\end{equation}

\noindent where $\hat{F}$ denotes the DFT Fock operator of a unit cell of
the quasi-1D crystal, and $F_{mn}^{\vec{k}}$ is its matrix element in Bloch
wave representation with $F_{mn}^{0,\vec{R}_{i}}$ standing for the coupling
Fock matrix element between lattice $0$ and $\vec{R}_{i}$ in real-space.
Theoretically, the summation in Eq.(\ref{Fmnk}) goes over all lattices in
real-space ($\vec{R}_{i}=0,\pm D,\pm 2D\cdots $). However, under TB
approximation, the summation in Eq.(\ref{Fmnk}) has only three terms
considered, including the on-site Hamiltonian $F^{0,0}$ and the nearest
neighbor coupling Hamiltonian $F^{0,1}$.

\begin{equation}
F_{mn}^{k}=F_{mn}^{0,0}+e^{ikD}F_{mn}^{0,D}+e^{-ikD}F_{nm}^{0,D}
\label{TBFmnk}
\end{equation}

And $S^{k}$ is calculated likewise

\begin{equation}
S_{mn}^{k}=S_{mn}^{0,0}+e^{ikD}S_{mn}^{0,D}+e^{-ikD}S_{nm}^{0,D}\text{.}
\label{TBSmnk}
\end{equation}%
\qquad\ \

Applying GAMESS(US) combined with NEGF subroutines, we first carry out the
DFT-NEGF calculation of the 4-ZGNR junction (Fig.\ref{alignment}) under zero
source-drain bias. Then, real-space matrices $F_{R}^{0,0}$, $S_{R}^{0,0}$, $%
F_{R}^{0,1}$, and $S_{R}^{0,1}$ are extracted from the converged result.
Substituting $F^{0,0}$, $S^{0,0}$, $F^{0,D}$, and $S^{0,D}$ in Eq.(\ref%
{TBFmnk}, \ref{TBSmnk}) with the extracted matrices respectively, TB band
structure $E(kD)$ of quasi-1D 4-ZGNR can be obtained by solving Roothaan
equation%
\begin{equation}
F^{k}C^{k}=S^{k}C^{k}E^{k}\text{.}  \label{Roothaan}
\end{equation}

We obtained a spin-polarized zero temperature ground state on 4-ZGNR with
spin dependent DFT-LSDA calculation, where the energy of the spin-polarized
state is about 20 meV per edge atom lower than the spin-unpolarized one,
similar to the results obtained by Son \emph{et al}.\cite{Young-Woo Son}
However, the magnetic order of ZGNR is reported to be unstable in the
presence of ballistic tunneling current under room temperature.\cite{Denis
A. Areshkin,symmetry,D.Gunlycke,H.Wagner,Z} Thus, we present a ZGNR device
simulation under room temperature with spin-unpolarized calculation in this
work.

\section{RESULTS AND DISCUSSIONS}

First, we apply NEGF-DFT calculation to the optimized structure of the
perfect quazi-1D 4-ZGNR, where two central unit cells including 32 carbon
atoms are treated as the scattering region between the contact layers and
periodic ZGNR electrodes (Fig.\ref{alignment}). PBC calculation of the
system is also performed by GAUSSIAN03 with exactly the same basis functions
and DFT calculation setup (CEP-4G and B3PW91) to check the validity of our
TB approximation.

\begin{figure}[tbp]
\includegraphics[width=1.0\textwidth]{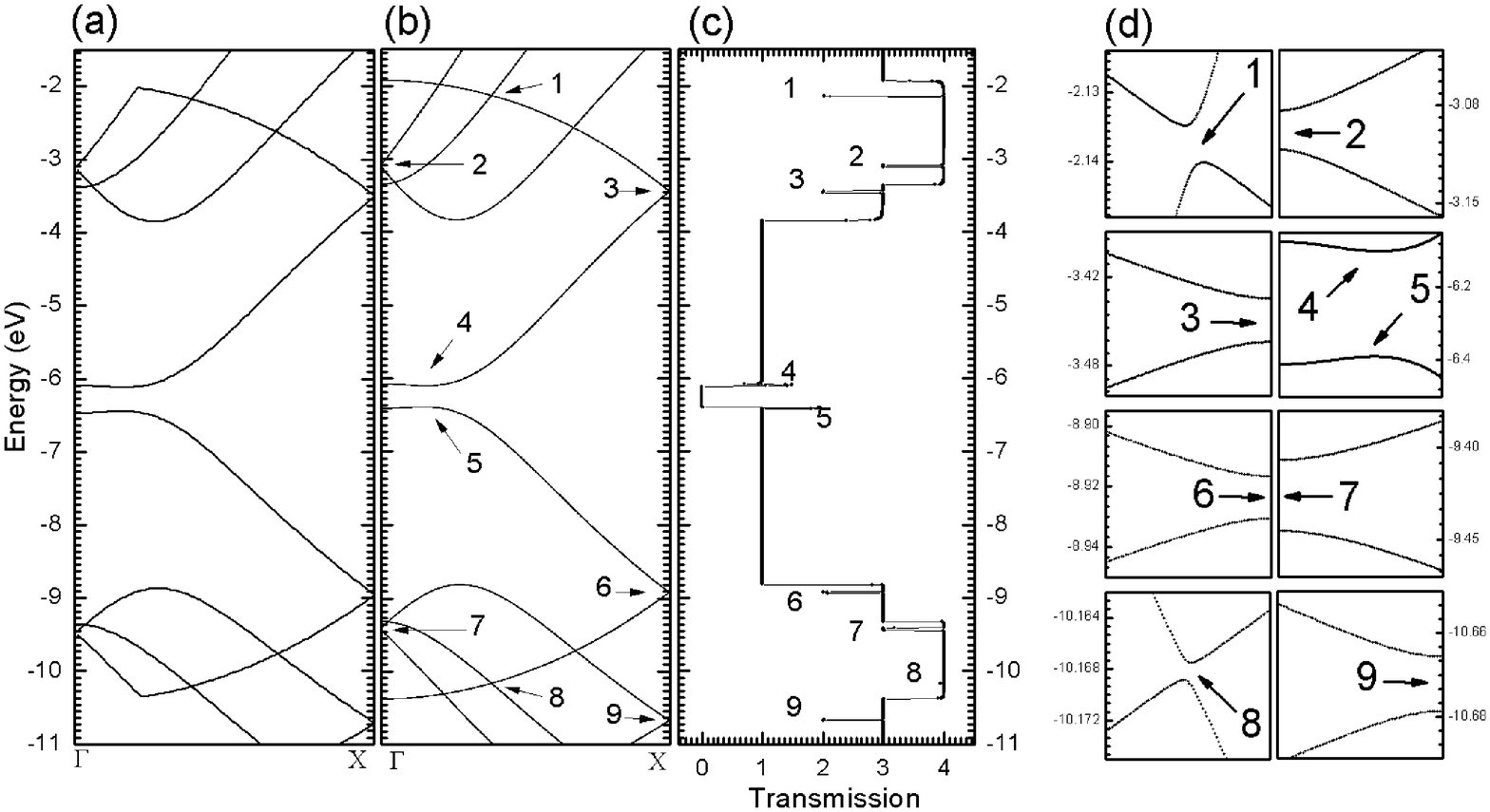}
\caption{G. Yin \emph{et al.}}
\label{Band}
\end{figure}

In Fig.\ref{Band}(a), the 4-ZGNR's PBC band structure provided by GAUSSIAN03
exhibits its semiconductor properties with a band gap of $0.3$ eV, while the
DFT-based TB band structure for the same system is presented in Fig.\ref%
{Band}(b) for comparison. Here, $E_{f}$ ($-6.2469$ eV) is at the center of
the band gap of 4-ZGNR. The result plotted is a so called "folded version"
because the unit cell is enlarged to include two armchair carbon layers.
Comparing Fig.\ref{Band}(a) and (b), the DFT-based TB band structure (b)
corresponds with the PBC result (a) perfectly, indicating that although only
$F_{mn}^{0,0}$ and $F_{mn}^{0,D}$ are included in the calculation, TB
approximation can still accurately predict the bulk property of 4-ZGNR:
almost all the interactions out of TB range are neglectable.

The semiconducting property of 4-ZGNR originates from two factors. In
addition to the transverse quantum confinement of the extremely narrow
4-ZGNR (about $1$ nm wide), the broken perfect honeycomb structure near the
GNR edges opens the gap on the band structure. As presented in Fig.\ref{unit
cell}, the edged C-C bond lengths of 4-ZGNR are no longer identical to the
ones in the perfect honeycomb structure, and the bond angles deviate from $%
120^{\circ }$ after structure relaxation. The reconfiguration of structure
breaks the perfect hexagon symmetry, which generates the band gap between $%
\pi $ and $\pi $* states. The semiconducting property of 4-ZGNR has its
disadvantage because of the poor conductance under low bias. However, the
C-C bonds between the leads and the scattering region are very strong, which
avoids the uncontrollable contact diversity on the interface between organic
molecules and metal leads. Our further calculation indicates that, with the
increase in the width of ZGNR, the gap gradually vanishes, and the unworking
region of ZGNR will be reduced. The width of GNR\ realized in experiments
are rarely smaller than $10$ nm nowadays, which is much wider than the
4-ZGNR considered in this work, making ZGNRs suitable to be applied as lead
material in realistic nanoscale circuits. However, the NEGF-DFT simulation
of a $10$ nm ZGNR is extremely time consuming and requires enormous
computational resources. In this work, we are not trying to reproduce an
actual experimental situation, but to demonstrate theoretically the
relationship between the band structure and the transmission spectrum for a
simple 4-ZGNR system, and to propose the mechanism that might control the
transport behavior of ZGNR devices.

The transmission spectrum $T\left( \epsilon \right) $ calculated under zero
bias is shown in Fig.\ref{Band}(c) with several integer steps and narrow
jump points, which is consistent with the band structures made by our
DFT-based TB (b) or g03 PBC (a) calculations. This is the usual
characteristics of the transmission spectrum for infinite lead. Similar
appearances were reported by other works for the study of quazi-1D systems.%
\cite{HP Cheng,symmetry,XG Zhang}

In such a perfect 1D crystal without any defects, each Bloch state can
tunnel through the junction without any classical resistance, and contribute
a unit of quantum conductance for the total transmission spectrum. The
integer steps in the transmission spectrum indicate the number of the
conducting channels or Bloch states. For the electrons with energy closely
higher than $-11$ eV, where cross three $E(kD)$ curves, three Bloch states
function as conducting channels, so the transmission spectrum is $3$ around
this energy value. In another case, the band structure $E(kD)$ crosses the
energy of $-10$ eV four times, so that the transmission here is $4$. In this
regime, for any energy value under consideration, one can accurately predict
the transmission coefficient for the perfect quasi-$1$D crystal simply by
counting the number of Bloch state in the electronic structure. The narrow
"jump points" in transmission spectrum (Fig.\ref{Band}(c)) are labeled from $%
1$ to $9$. They are not caused by numerical inaccuracy, but induced by the
sudden change in the number of the conduction channels. In Fig.\ref{Band}(d)
we enlarge the band structure around the points to show that there are
mini-gaps rather than band crossings corresponding to the transmission
abnormal points. Investigating carefully, one may find that the number of
Bloch states changes from one to two then zero when the relevant energy scan
up across the maximum of the band then entering the mini-gap (point $5$),
which causes the transmission value varied from one to two then zero. The
sudden vanishment of two Bloch states inside the narrow band gap of $0.0056$
eV explains the sharp drop of two quantum conductance units in the
transmission function at point $1$. The jump points $2$, $3$, $6$, $7$ and $%
9 $ are all companied by the small band gaps on $\Gamma $ or X points,
generating a decline of one quantum conductance unit in the transmission
function at the relevant energy values due to the absence of one Bloch
state. Interestingly, the runtish point $8$ fails to achieve an integer
drop. Its exotic behavior is caused by the tiny mini-gap of $0.0013$ eV,
which is very hard to demonstrate by the transmission function with the
numerical sampling step length as large as $0.005$ eV. Fortunately,
tunneling current is evaluated by the integral of transmission spectrum, and
the numerical error like point $8$ does not influence the result
significantly even if the error is included in the bias window. It should be
pointed out that although not shown here, all other crossings in the TB band
structure are also investigated in details, with $10^{4}$ sampling k points
from $\Gamma $ to X. No band gaps can be observed at the energy values where
no abnormality exists on transmission function, demonstrating a very
accurate matching between DFT-NEGF calculated transmission and the DFT-TB
band structure. This matching assures the validity of the SGF of left
(right) electrode, which is obtained by the DFT-TB calculation.

In order to investigate the external electrostatic field response in 4-ZGNR,
homogeneous transverse electric fields are applied to the scattering region.
The orientation of external field is arranged perpendicularly to the GNR
axis (shown in Fig.\ref{Pure}), with the intensity of $1.028$ V/\AA\ ($0.02 $
a.u.). By the help of MacMolPlt,\cite{Bode} 3D HOMO LUMO and 2D molecular
electrostatic potential (MEP) distribution of pure 4-ZGNR are presented in
Fig.\ref{Pure} with the 2D plotting major plane parallel to the ZGNR and $%
2.0 $ \AA\ away from it. The flat surface of 2D MEP map is specifically
chosen to reveal the electrostatic potential contributed by the $\pi $ and $%
\pi ^{\ast }$ hybridized orbitals, which are mainly formed by the overlap of
$p_{x}$ orbital of carbon atoms. The red contours are positive potential
isolines, while the blue ones are negtive. We note that the MEP plot is
generated based on the optimized orbitals and occupations computed by GAMESS
in the presence of the external field, subttracting the contribution of the
external field itself. Similar to previous works, HOMO and LUMO are equally
distributed on edges, and the magnitude of electron wave function decays
from edge to the center gradually.\cite{Young-Woo Son} Under external
fields, the molecular orbitals of $\pi $ and $\pi ^{\ast }$ are splitted as
the consequence of electric polarization: HOMO concentrated at the
low-potential edge, while LUMO at the high-potental edge. Unfortunately, the
transmission function of pure ZGNR (not shown here) hardly responds to
external fields: the fields can only control the orbital coupling in the
direction perpendicular to the tunneling current.

In recent first-principle calculations, some groups indicate that vacancies
on GNR edges are energetically preferred\cite{DFCT Structure}, which
significantly suppresses GNRs' conductance.\cite{defect NEGF} Here, we
fabricate GNR edge defects by removing three carbon atoms from the edges of
4-ZGNR, with the dangling bonds saturated by hydrogen atoms. The central
junction structure including 58 carbon atoms is optimized with the left and
right leads fixed in their PBC optimized positions. The contact layers are
kept out of the TB range from the central defect region, so that their
"bulk" properties are protected by the buffer layers. Optimization result
indicates that, all carbon and hydrogen atoms in the defect region are still
in the same plane after relaxation, and the saturating hydrogen atoms pushes
the edge carbon atoms off the vacancy significantly. The equilibrium
transmission spectrum of the edge defect junction is presented in Fig.\ref%
{TF} with the external field $E_{ex}$ adjusted from $0$ to $0.771$ V/\AA\ ($%
0.015$ a.u.). Polarization effect stimulated by the transverse external
fields for the HOMO and LUMO distribution is shown in Fig.\ref{DFCT PLRZTN}
respectively. Similar to reported results, the GNR defect junction exhibits
about 50\% decline of transmission around $E_{f}$ under zero external
fields. This drastic conductance depression is obvious because pure ZGNR's
HOMO and LUMO are mainly distributed on edges so that their orbital coupling
in the tunneling direction is significantly suppressed with edge defects
presented. As plotted in Fig.\ref{DFCT PLRZTN}, edge vacancies break two of
the six-sided carbon rings, and the C-C \emph{sp}$^{2}$ hybridization of $%
\pi $ and $\pi ^{\ast }$ MO is substituted by the localized C-H bonds on the
edge vacancy, leading to a radical disruption of the delocalized conjugated
system.

Although the conductance of defect junction is no better than pure ZGNR, the
presence of edge defects makes it more sensitive to the modulation of
external transverse electric fields, which leads the possible application of
the defect ZGNR junctions to nanoscale electronic engineering. The I-V curve
of the defect junction given by Eq.(\ref{cur}) is presented in Fig.\ref{I-V}%
, in which a very clear depression of tunneling current can be observed
under large transverse external fields. In this calculation, the temperature
of Fermi distribution is set at room temperature ($T=300$K). The low
conductance under the voltage $V_{SD}<0.3$ V is caused by the HOMO-LUMO gap
of 4-ZGNR. The tunneling current increases considerably with applied bias
exceeding the gap threshold. In the presence of external fields, the
junction almost turns off the conducting channel when $E_{ex}$ exceeds $1.0$
V/\AA , exhibiting the maximum of the on/off ratio up to $10^{4}$. Similar
with the situation of pure ZGNR, LUMO and HOMO of the defect junction are no
longer symmetrically distributed when $E_{ex}$ is applied. HOMO (electron
states) and LUMO (hole states) are polarized to opposite edges.(Fig.\ref%
{DFCT PLRZTN}) Contrasting to pure 4-ZGNR, electric polarization depresses
the local orbital coupling of the central region to the GNR electrode much
more severely as a consequence of the edge vacancies, producing significant
difficulties for electrons to tunnel through the junction. To present a
clearer mechanism of the conductance decay induced by electric polarization,
MEP map of the junction's scattering region is plotted in Fig.\ref{Isoline}
with the potential of external field subtracted off. The isoline circles in
the MEP indicate charge concentration induced by polarization, with positive
charge conglomerated on the upper edge, and the negative one on the
opposite. Functioning together with the geometry shape, tunable external
fields splits the electron wave function with a potential barrier
constructed on the edge vacancies, reflecting the carriers back to the
electrodes. Since no structural modification is required in the control
process, conductance manipulation based on electric polarization is more
sensitive to external stimulus, so GNR edge defect junctions have the
potential to be applied as switching devices with considerable responding
frequencies under the control of transverse electric fields. On the other
hand, for a pure 4-ZGNR without any vacancy, although polarization exits,
the potential barrier can not be established, hence the tunneling current
barely responds to $E_{ex}$ sensitively.

\section{SUMMARY}

In summary, we use DFT-NEGF method to simulate the transport properties of
4-ZGNR, and find out that edge defects make the tunneling current of ZGNR
junctions more controllable by the transverse external electrostatic fields.
In this work, ZGNR electrodes are modeled with DFT-based TB approximation,
and the validity of the approximation is assured by comparison between the
perfect ZGNR band structures calculated under our DFT-based TB and
GAUSSIAN03 PBC calculations. Using DFT-based TB method, we obtain the
transmission function of the perfect ZGNR system, of which the integer steps
and jump points are explained well by the band structure. Electronic devices
made from the edge defect ZGNR junctions have the advantage of easy control
through the transverse electrostatic fields. The study of small and narrow
GNR devices benefits graphene integrated circuit engineering that might be
realized by ultra fine GNR fabrication technologies in the future.

\begin{acknowledgements}
We feel grateful for constructive and inspiring discussions between
one of us (G.Y.) and Prof. Xiao-Guang Zhang at ORNL. This work is
also sponsored by Natural Science Foundation of China (NSFC) under
Grant No. 90606024, the Program for Major Research Plan of China
(Project No. 2006CB 921302), and Nippon TechnoLab Co. Ltd. We thank
the support from Fudan High-end Computing Center, Shanghai
Supercomputer Center, and SR11000 supercomputer from the Center for
Computational Materials Science of Institute for Materials Research,
Tohoku University.
\end{acknowledgements}

\newpage

\begin{flushleft}
{\large \textbf{Captions} }
\end{flushleft}

\begin{description}
\item {FIG. 1.} (Color online) The arrangement of 4-ZGNR based junction
after geometry optimization with dangling bonds on zigzag edges saturated by
hydrogen atoms. Two armchair carbon layers are included in each TB unit cell
to avoid sever inter-cell interaction loss of TB approximation. In DFT-NEGF
calculation for a pure 4-ZGNR, two TB unit cells are treated as the
scattering region between the left and right contact layers.

\item {FIG. 2.} Band structure of perfect 4-ZGNR quasi-$1$D crystal drawn
under (a) periodic boundary condition and (b) tight-binding approximation.
Only 8 energy bands around the Fermi level are presented in the PBC band
structure. Transmission spectrum of the scattering region (c) is presented
with 9 narrow transmission jump points. Detailed band structures around
these points are demonstrated in (d).

\item {FIG. 3.} (Color online) Fully relaxed unit cell alignment of 4-ZGNR:\
the perfect graphene honeycomb structure is broken, which is one of the
reasons of 4-ZGNR's semiconducting property. Length unit in this plot is \AA %
.%

\item {FIG. 4.} LUMO (a)(d) and HOMO (b)(e) distribution on pure 4-ZGNR.
Positive (red) and negative (blue) 2D molecular electrostatic potential
(MEP) isolines of pure ribbon are shown in (c) and (f). The arrows in (d),
(e), and (f) indicate the transverse external field. 

\item {FIG. 5.} Zero bias transmission spectrum of edge defect junction.

\item {FIG. 6.} LUMO and HOMO of edge defect junction, with $E_{ex}$
indicated by blue arrows. 

\item {FIG. 7.} \emph{I-V} curves of edge defect junction in modulation of
the intensity of transverse electric field. The inset shows the optimized
structure of edge defect junction and the setup of the scattering region.
The environment temperature of this simulation is $300$ K.%
%
%
%
%
%
%
%
%
%
%

\item {FIG. 8.} 2D electrostatic potential distribution of edge defect
junction. The plotting plane is parallel to the ribbon, and 1.2 \AA\ away
from it. 
\end{description}

\clearpage\newpage

\begin{figure}[tbp]
\includegraphics[width=0.4\textwidth]{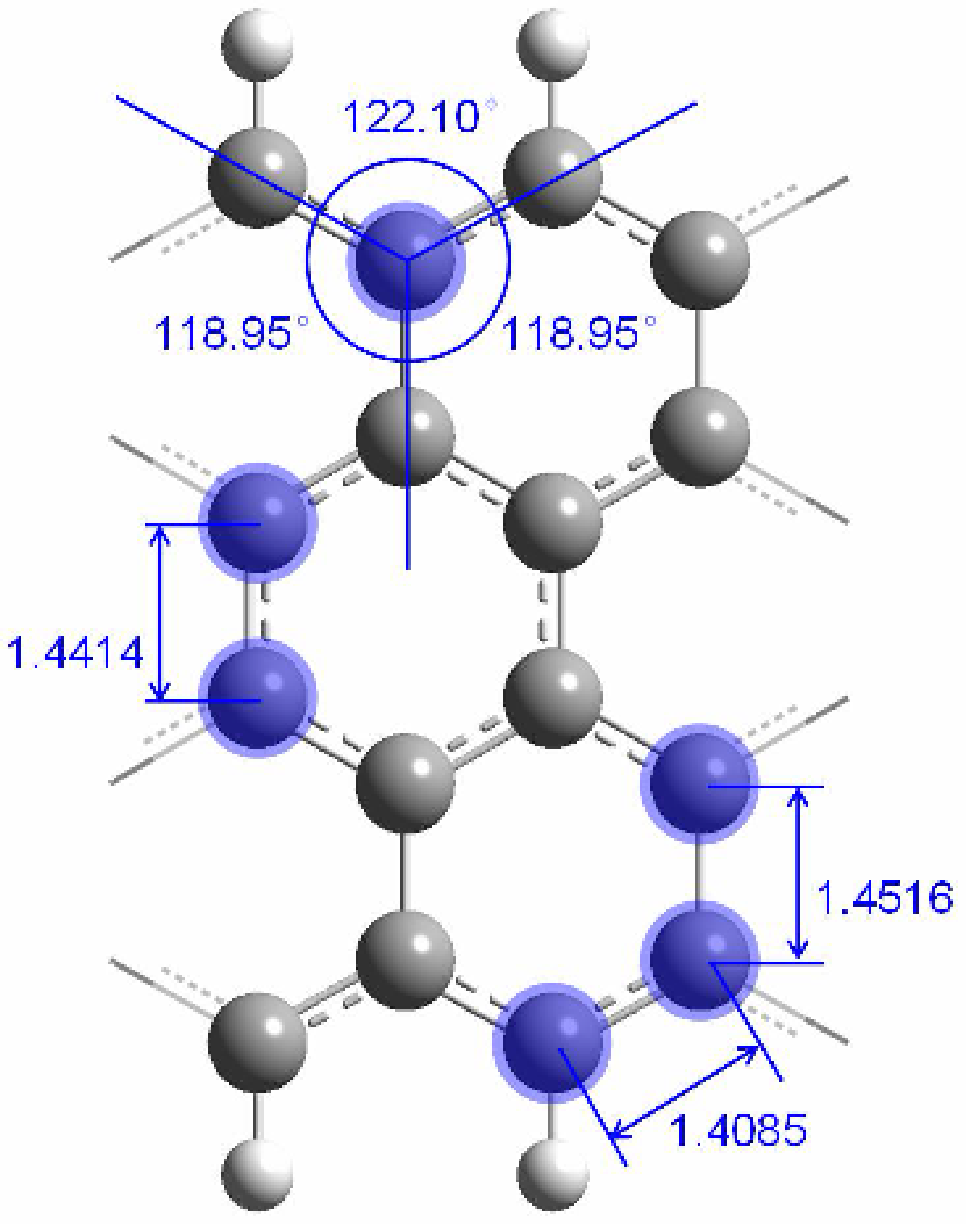}
\caption{G. Yin \emph{et al.}}
\label{unit cell}
\end{figure}

\begin{figure}[tbp]
\includegraphics[width=1.0\textwidth]{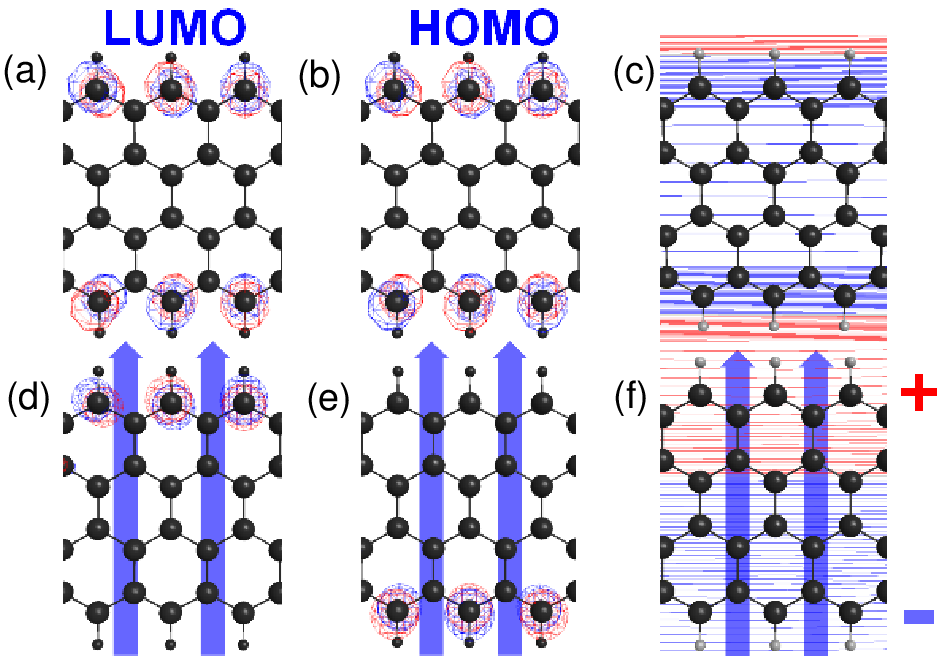}
\caption{G. Yin \emph{et al.}}
\label{Pure}
\end{figure}

\begin{figure}[tbp]
\includegraphics[width=1.0\textwidth]{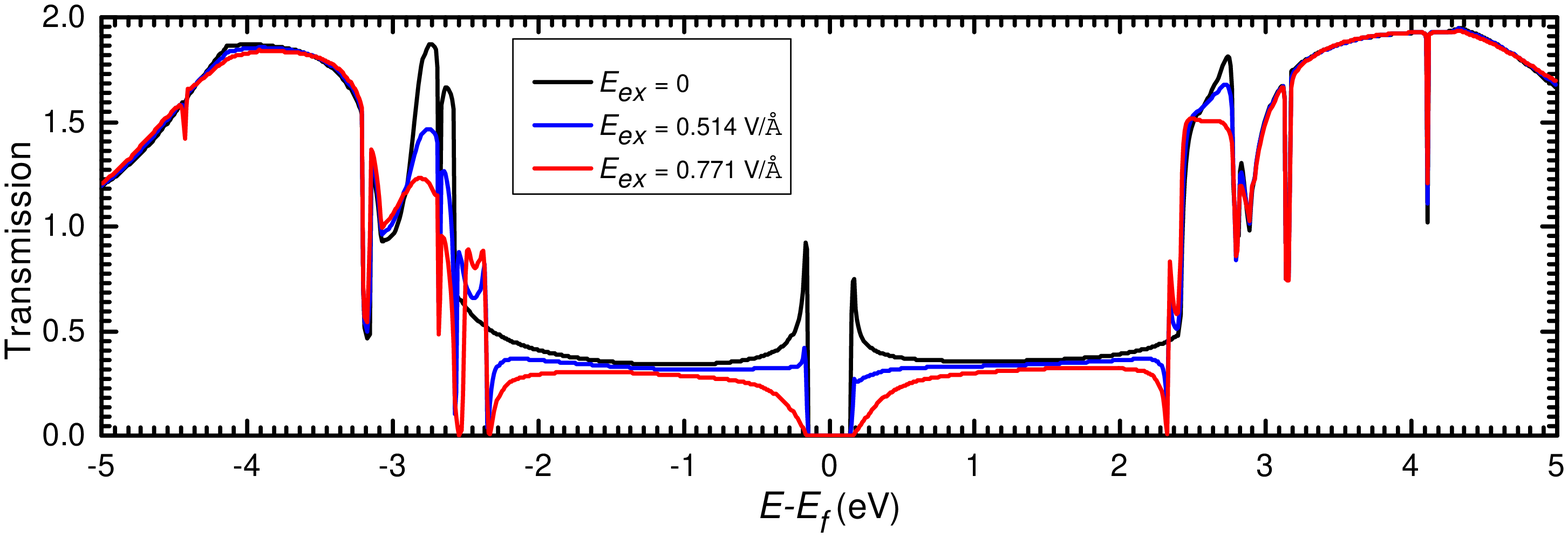}
\caption{G. Yin \emph{et al.}}
\label{TF}
\end{figure}

\begin{figure}[tbp]
\includegraphics[width=1.0\textwidth]{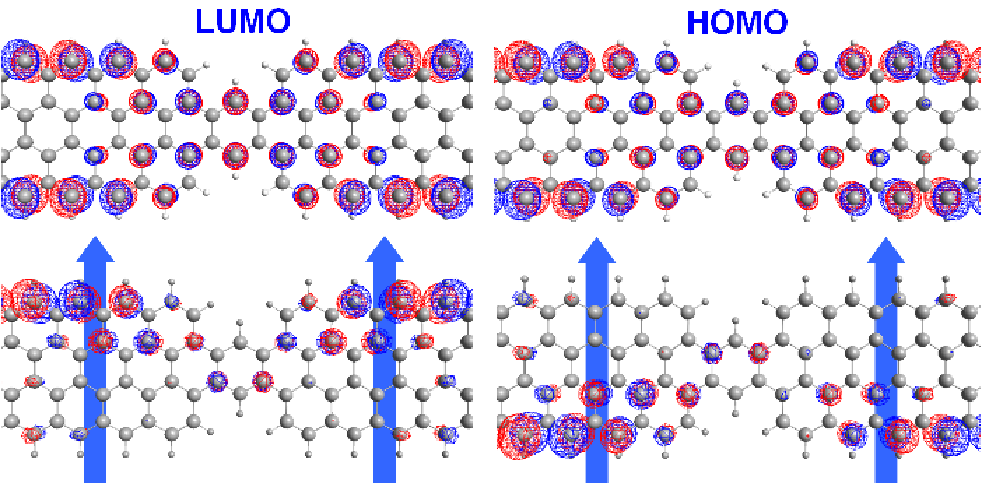}
\caption{G. Yin \emph{et al.}}
\label{DFCT PLRZTN}
\end{figure}

\begin{figure}[tbp]
\includegraphics[width=1.0\textwidth]{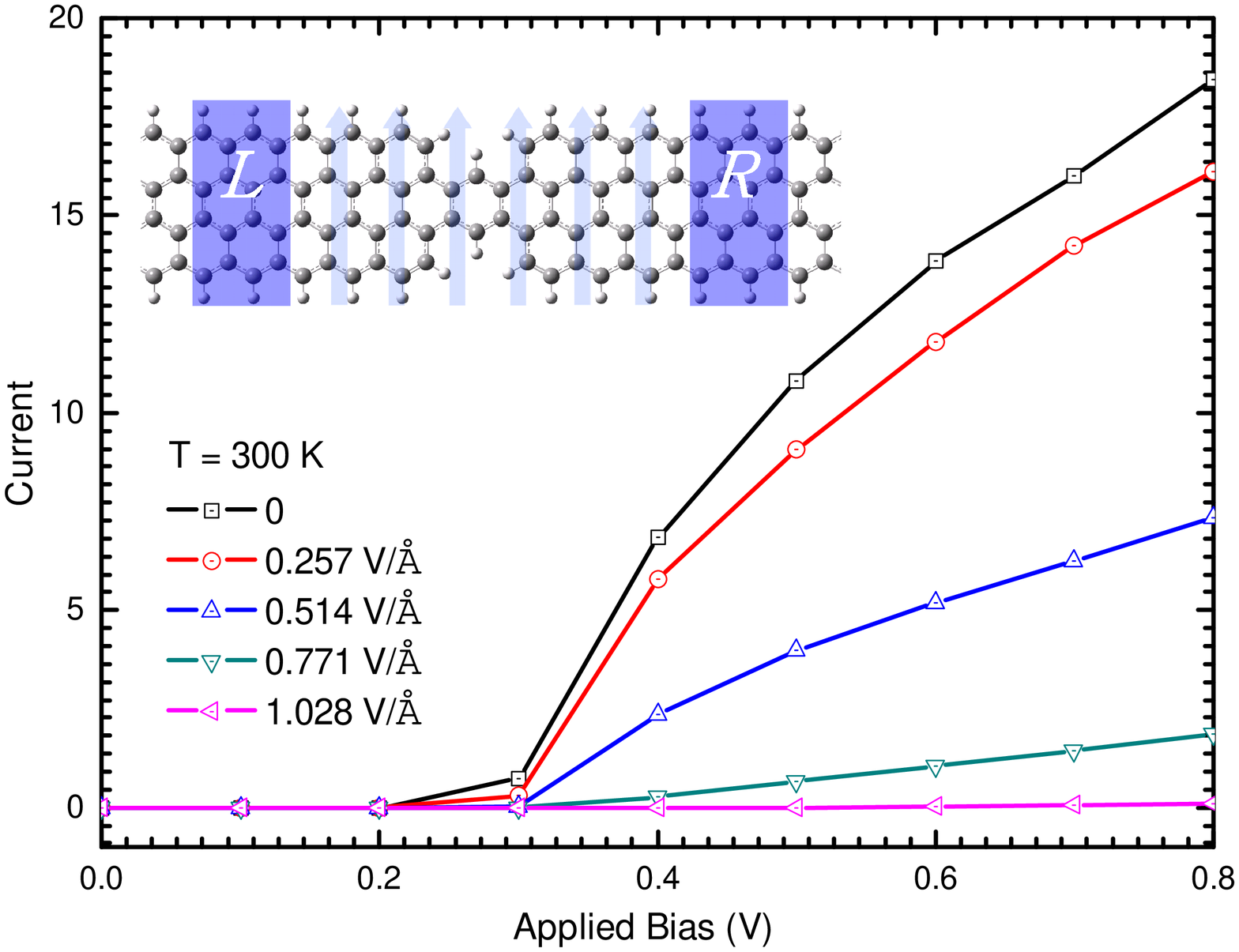}
\caption{G. Yin \emph{et al.}}
\label{I-V}
\end{figure}

\begin{figure}[tbp]
\includegraphics[width=1.0\textwidth]{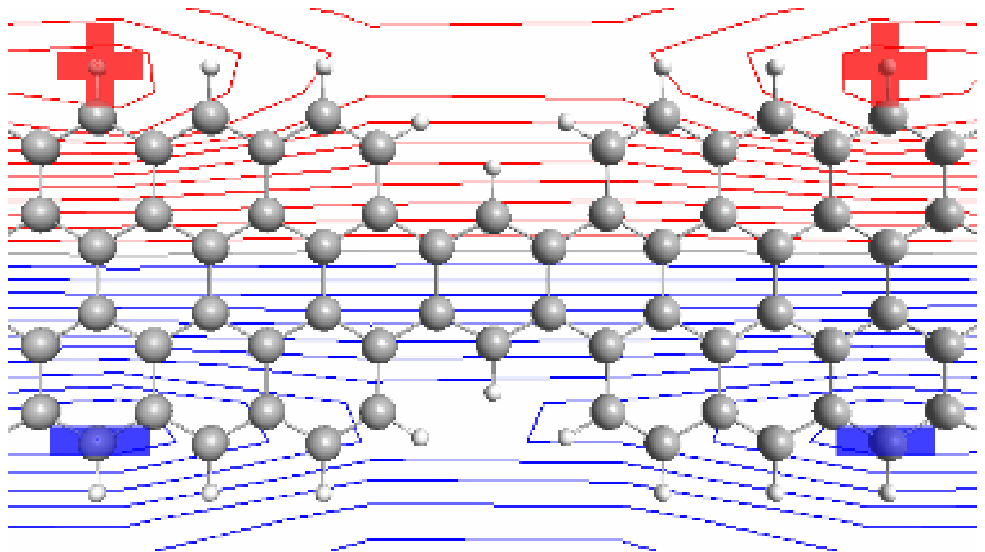}
\caption{G. Yin \emph{et al.}}
\label{Isoline}
\end{figure}

\end{document}